\documentclass[letterpaper]{aipproc}
\layoutstyle{8x11single}

\SetInternalRegister\hbadness{8000} 

\newcommand\doingARLO[2][]{%
  \ifx\mmref\undefined #1\else #2\fi
}

\begin{document}

\title 
      [New Results on Pb-Au Collisions at 40 AGeV from the CERES/NA45 Experiment]
      {New Results on Pb-Au Collisions at 40 AGeV from the CERES/NA45 Experiment}


\author{K. Filimonov$^1$\footnote{Present address: LBNL, Berkeley, USA}}{
  address={Universit\"at Heidelberg, Germany}
}
\author{D.~Adamov\'a}{
  address={NPI/ASCR, \v{R}e\v{z}, Czech Republic}
}
\author{G.~Agakichiev}{
  address={GSI, Darmstadt, Germany}
}
\author{H.~Appelsh\"auser}{
  address={Universit\"at Heidelberg, Germany}
}
\author{V.~Belaga}{
  address={JINR, Dubna, Russia}}
\author{P.~Braun-Munzinger}{
  address={GSI, Darmstadt, Germany}
}
\author{A.~Cherlin}{
  address={Weizmann Institute, Rehovot, Israel}
}
\author{S.~Damjanovi\'c}{
  address={Universit\"at Heidelberg, Germany}
}
\author{T.~Dietel}{
  address={Universit\"at Heidelberg, Germany}
}
\author{A.~Drees}{
  address={Department for Physics and Astronomy, SUNY Stony Brook, USA}
}
\author{S.\,I.~Esumi}{
  address={Universit\"at Heidelberg, Germany}
}
\author{K.~Fomenko}{
  address={JINR, Dubna, Russia}
}
\author{Z.~Fraenkel}{
  address={Weizmann Institute, Rehovot, Israel}
}

\author{C.~Garabatos}{
  address={GSI, Darmstadt, Germany}
}
\author{P.~Gl\"assel}{
  address={Universit\"at Heidelberg, Germany}
}
\author{G.~Hering}{
  address={GSI, Darmstadt, Germany}
}
\author{V.~Kushpil}{
  address={NPI/ASCR, \v{R}e\v{z}, Czech Republic}
}
\author{B.~Lenkeit}{
  address={CERN, Geneva, Switzerland}
}
\author{A.~Maas}{
  address={GSI, Darmstadt, Germany}
}
\author{A.~Mar\'{\i}n}{
  address={GSI, Darmstadt, Germany}
}
\author{J.~Milo\v{s}evi\'c}{
  address={Universit\"at Heidelberg, Germany}
}
\author{A.~Milov}{
  address={Weizmann Institute, Rehovot, Israel}
}
\author{D.~Mi\'skowiec}{
  address={GSI, Darmstadt, Germany}
}
\author{Y.~Panebrattsev}{
  address={JINR, Dubna, Russia}
}
\author{O.~Petchenova}{
  address={JINR, Dubna, Russia}
}
\author{V.~Petra\v{c}ek}{
  address={Universit\"at Heidelberg, Germany}
}
\author{A.~Pfeiffer}{
  address={CERN, Geneva, Switzerland}
}
\author{J.~Rak}{
  address={GSI, Darmstadt, Germany}
}
\author{I.~Ravinovich}{
  address={Weizmann Institute, Rehovot, Israel}
}
\author{P.~Rehak}{
  address={Brookhaven National Laboratory, Upton, USA}
}
\author{H.~Sako}{
  address={GSI, Darmstadt, Germany}
}
\author{W.~Schmitz}{
  address={Universit\"at Heidelberg, Germany}
}
\author{J.~Schukraft}{
  address={CERN, Geneva, Switzerland}
}
\author{S.~Sedykh}{
  address={GSI, Darmstadt, Germany}
}
\author{W.~Seipp}{
  address={Universit\"at Heidelberg, Germany}
}
\author{S.~Shimansky}{
  address={JINR, Dubna, Russia}
}
\author{J.~Sl\'{\i}vov\'a}{
  address={Universit\"at Heidelberg, Germany}
}
\author{H.\,J.~Specht}{
  address={Universit\"at Heidelberg, Germany}
}
\author{J.~Stachel}{
  address={Universit\"at Heidelberg, Germany}
}
\author{M.~\v{S}umbera}{
  address={NPI/ASCR, \v{R}e\v{z}, Czech Republic}
}
\author{H.~Tilsner}{
  address={Universit\"at Heidelberg, Germany}
}
\author{I.~Tserruya}{
  address={Weizmann Institute, Rehovot, Israel}
}
\author{J.\,P.~Wessels}{
  address={Universit\"at Heidelberg, Germany}
}
\author{T.~Wienold}{
  address={Universit\"at Heidelberg, Germany}
}
\author{B.~Windelband}{
  address={Universit\"at Heidelberg, Germany}
}
\author{J.\thinspace P.~Wurm}{
  address={Max-Planck-Institut f\"ur Kernphysik, Heidelberg, Germany}
}
\author{W.~Xie}{
  address={Weizmann Institute, Rehovot, Israel}
}
\author{S.~Yurevich}{
  address={Universit\"at Heidelberg, Germany}
}
\author{V.~Yurevich}{
  address={JINR, Dubna, Russia}
}

\copyrightyear  {2001}

\date{\today}

\maketitle

\section{Introduction}
Low-mass dilepton spectra are expected to provide information on the 
early stage of the relativistic heavy-ion 
collisions which may experience an onset of deconfinement and/or 
chiral symmetry restoration. 
The CERES/NA45 measurements of $e^+e^-$-pair production in 158 AGeV Pb-Au 
collisions revealed a significant excess of the dielectron yield over 
known hadronic sources in the 
invariant mass region 0.25<m$_{ee}<0.7$ GeV/c$^2$ \cite{bjoernqm}.
The enhancement is mostly observed for pairs with low transverse momentum
$p_t^{ee}<0.5$ GeV/c and increases stronger than linearly with the event multiplicity. 
These experimental data have stimulated extensive theoretical
discussions on the mechanisms of low-mass dilepton production and emission from
hot and dense hadronic medium (for a review see \cite{review}). The thermal 
radiation from the interaction phase of the hadronic fireball via annihilation
processes, predominantly $\pi^+\pi^-\rightarrow\rho\rightarrow e^+e^-$, may account for a large fraction of the observed enhancement. The spectral shape in the mass region of the excess, 
however, cannot be described and requires introducing in-medium effects on the $\rho$ spectral function. Two 
scenarios that take into account in-medium modifications of the vector meson properties 
have been particularly successful in reproducing the CERES results. One 
assumes a reduction of the $\rho$-meson mass in
the hot and dense medium as a precursor of chiral symmetry restoration \cite{dropping}, following Brown-Rho scaling \cite{brownrho}. Another
approach uses a $\rho$-meson spectral function which takes into account 
modifications of meson properties due to interactions with the surrounding 
hadrons \cite{broadening}. Dilepton production rates and shapes of the invariant mass spectra
in these two calculations
are very similar pointing to a shift of the hadron-parton duality threshold towards lower masses \cite{duality}. 
The measurement of dilepton spectra at lower beam energy of 40 AGeV allows to
study the effect of varying the baryon density and provides additional 
constraints for the model calculations. 
It is also the first measurement after
upgrading the CERES spectrometer with a Time Projection Chamber to provide a 
better resolution in the mass region of the narrow resonances $\omega$ 
and $\phi$. 

\section{Experimental Setup}

\begin{figure}[htb]
  \resizebox{0.7\textwidth}{!}{\includegraphics[angle=270]{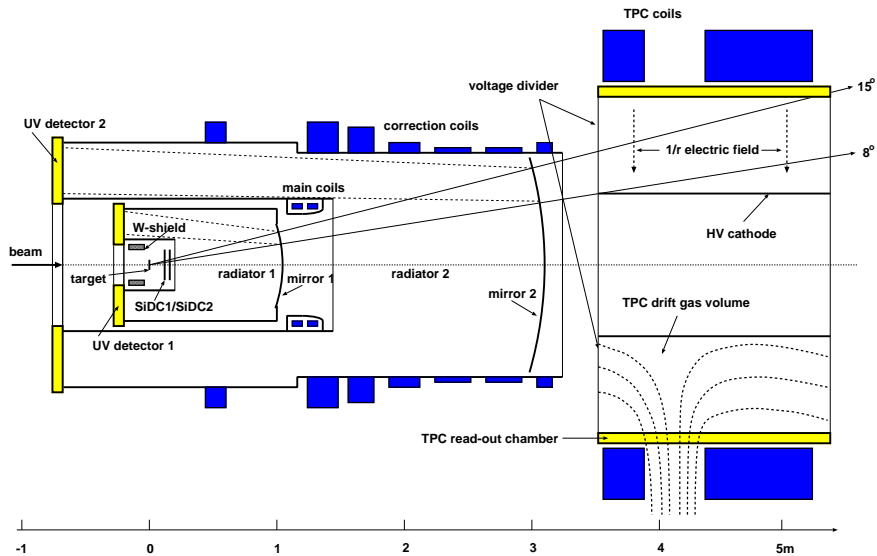}}
  \caption{The upgraded CERES spectrometer at the CERN-SPS}
\label{fig:setup}
\end{figure}

The CERES spectrometer (Figure \ref{fig:setup}) is optimized to measure electron pairs at midrapidity 
($2.1<\eta<2.6$) with full azimuthal coverage. Two silicon drift chambers 
(SIDC1,2) located 10 cm and 13.8 cm behind a segmented Au target provide a 
precise angular measurement of charged particles and vertex reconstruction.
The electrons are identified by the two Ring Imaging Cherenkov detectors
(RICH1,2) operated at a $\gamma_{\rm th}=32$, which rejects 95\% of 
all charged 
hadrons. The momentum of all charged particles is measured by the radial drift
Time Projection Chamber (TPC) \cite{anaqm}. The TPC has an active length of 2 m and an outer
diameter of 2.6 m and provides the measurement of 20 space points in a magnetic
field with a maximal radial component of 0.5 T. 

The data taking period in the fall of 1999 was the first run of CERES after the
TPC upgrade. Due to problems with the new readout the data set is 
limited in statistics and TPC efficiency. In total we recorded 8.7M Pb-Au 
events at 40 AGeV with centrality $\sigma/\sigma_{\rm geo}\approx 30$\%.

\section{Electron-Positron Pair Analysis}

The dominant sources of $e^+e^-$-pairs are photon conversions and 
$\pi^0$-Dalitz decays which are characterized by the small pair opening angle
and invariant masses below 200 MeV/c$^2$. The large number of such pairs 
(N$_{e^+e^-}$(m<200 MeV/c$^2$)/N$_{e^+e^-}$(m>200 MeV/c$^2$)$\sim 10^4$) 
combined with  
limited track reconstruction efficiency and acceptance results in 
a combinatorial background for events with two partially reconstructed low-mass
pairs. 
This combinatorial background can be significantly 
reduced by pairing only tracks with transverse momentum $p_t>200$~MeV/c.
To remove the hadronic contamination in accidental matching between RICH and TPC we use
d$E$/d$x$ information from the TPC. Conversions and Dalitz
pairs with opening angle less than 10 mrad which are not recognized as
two individual rings in the RICH detectors are rejected by a double energy loss 
cut in the two SIDC detectors. To account for a limited TPC efficiency in 
the 1999 data set, we also remove tracks which have a SIDC-RICH electron
candidate within 70 mrad. Finally, identified Dalitz pairs (m$_{ee}<200$ MeV/c$^2$) are excluded from further combinatorics. The signal is extracted by subtracting like-sign pairs from unlike-sign pairs. The total number of open pairs (m$_{ee}>200$ MeV/c$^2$) is $180\pm48$ (stat.) with a signal-to-background ratio
of 1/6. 

\begin{figure}[htb]
  \resizebox{.5\textwidth}{!}{\includegraphics{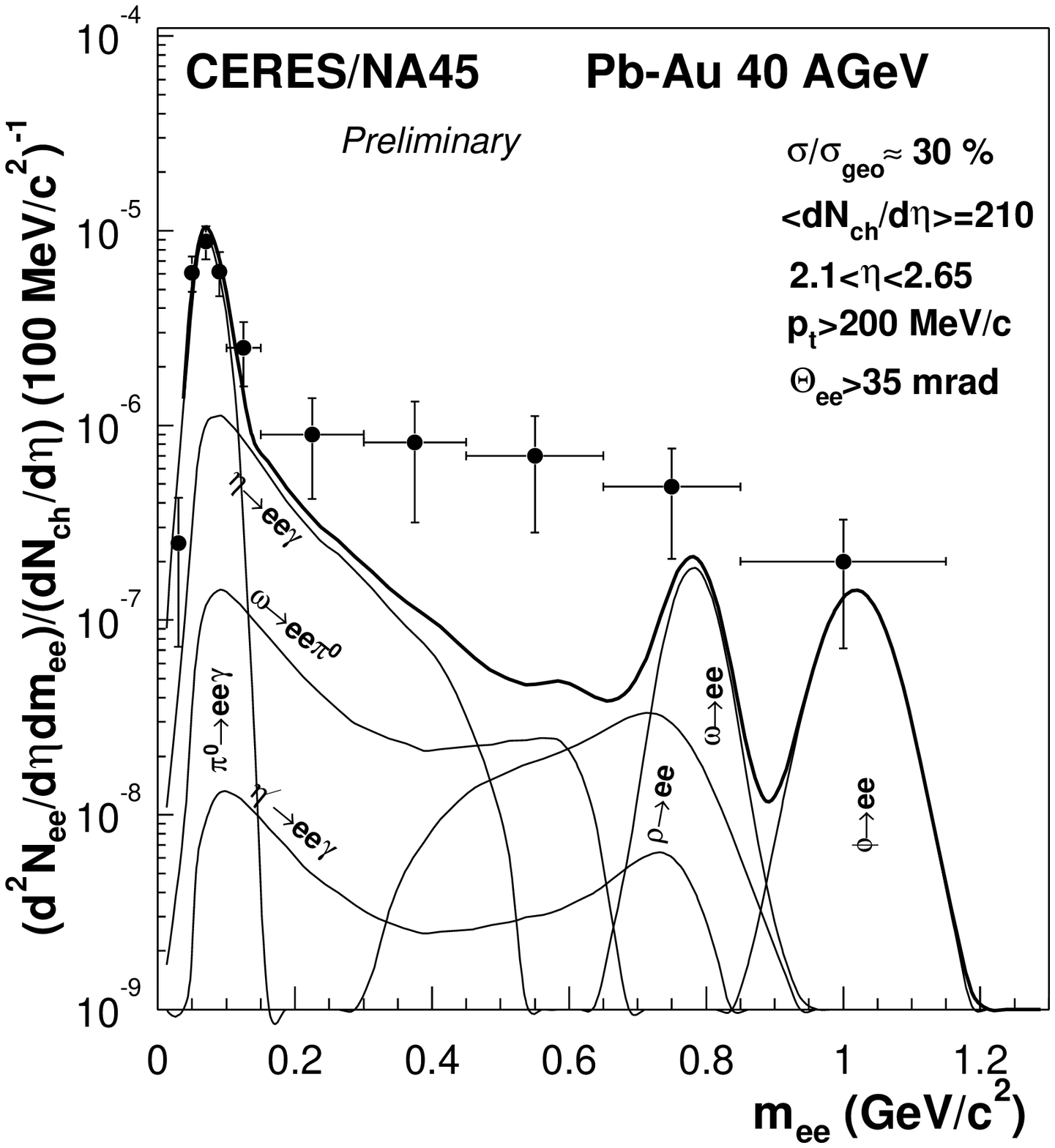}}
  \resizebox{.5\textwidth}{!}{\includegraphics{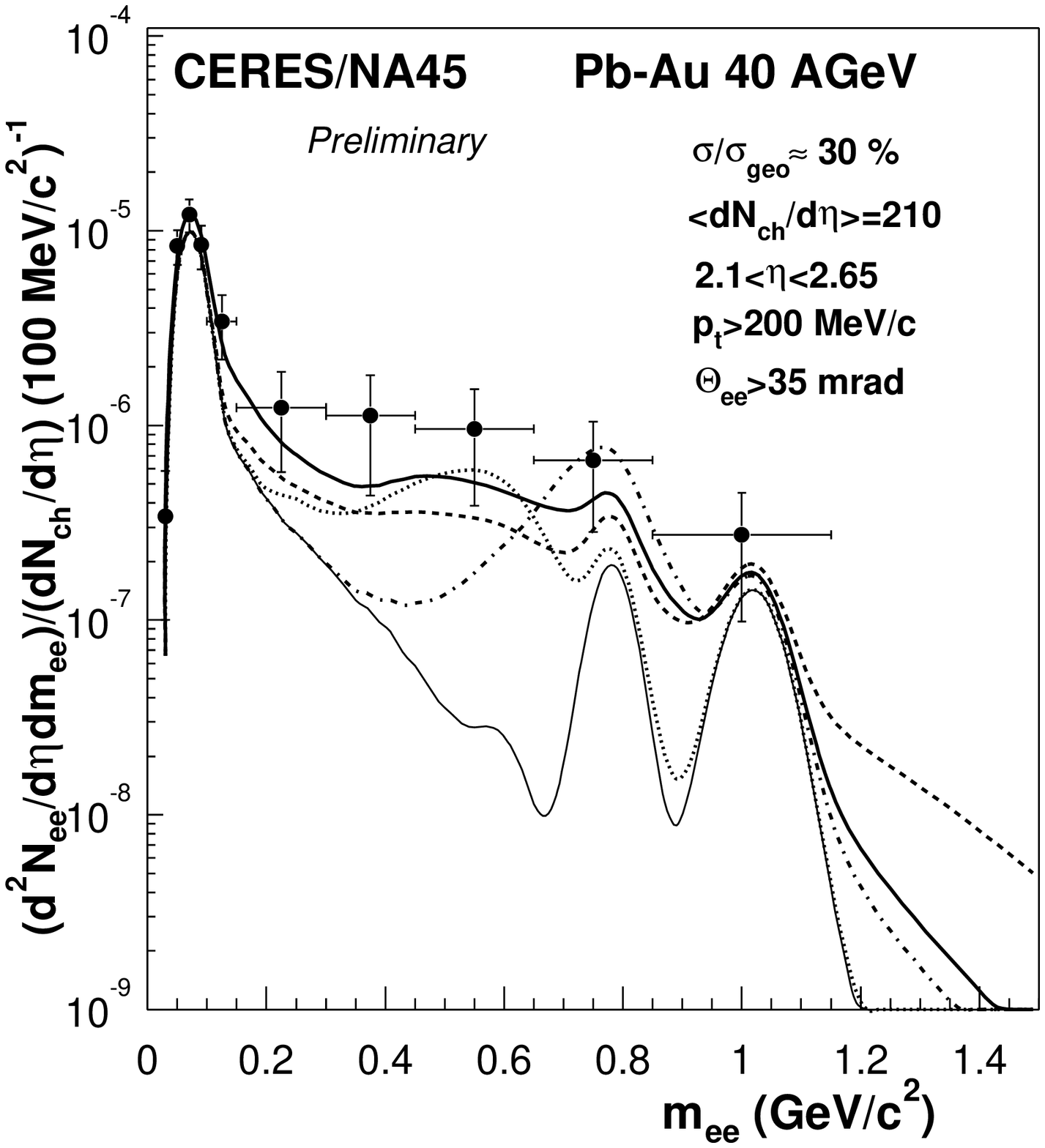}}
  \caption{Inclusive $e^+e^-$ invariant mass spectrum at 40 AGeV
in comparison to expectations from known hadronic sources 
in heavy ion collisions (`hadronic cocktail') (left) 
and in comparison to model calculations (right) assuming the vacuum 
$\rho$ spectral function (dash-dotted), a dropping $\rho$ mass (dotted)
and including medium modifications of the $\rho$ spectral function
(thick solid line). Also shown is the hadronic cocktail
without $\rho$ (thin solid line).
The dashed line refers to a lowest order pQCD rate calculation. 
All calculations
are from~\cite{rapppriv}.}
\label{fig:1-2}
\end{figure}

The measured $e^+e^-$ invariant mass spectrum is compared to the hadronic
cocktail in Figure~\ref{fig:1-2} (left). 
\vspace{1.0mm}
The data were normalized to the hadronic cocktail in the $\pi^0$-Dalitz 
region. Similar to the observations at 158 AGeV, the measured $e^+e^-$ yield exhibits an enhancement by a factor of $5.0\pm 1.5$ (stat.) in the mass 
range m$_{ee}$>200 MeV/c$^2$ compared to the expectations from known hadronic
sources in heavy-ion collisions. The excess is most pronounced for pairs with small transverse momentum. The results are also compared to different
theoretical calculations in Figure~\ref{fig:1-2} (right). The measurements 
clearly disfavor a purely hadronic scenario assuming the vacuum $\rho$ spectral function.

\section{Hadron Spectra and Yields}

\begin{figure}[htb]
\includegraphics[width=0.5\textwidth]{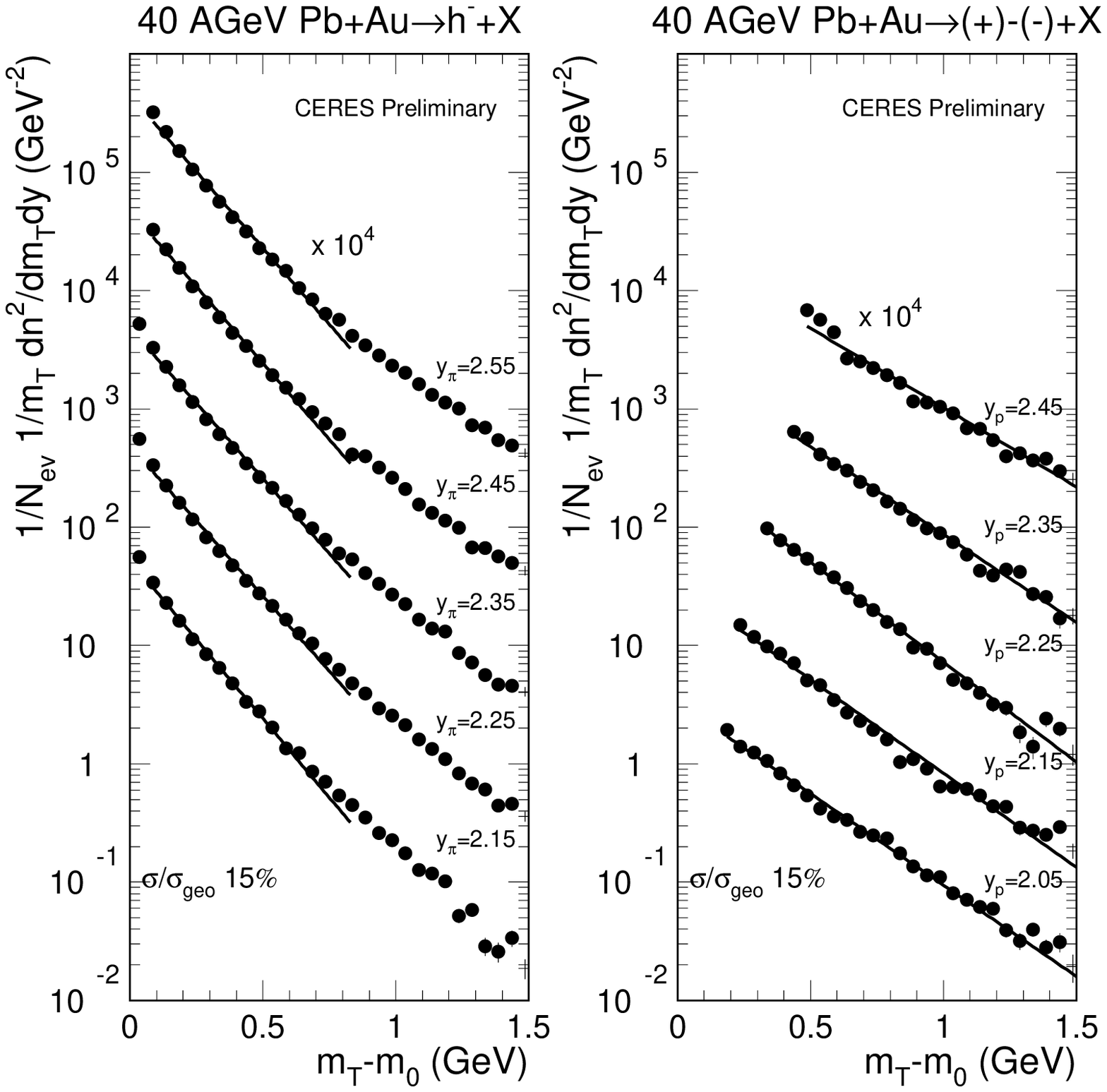}
\vspace{-1.5cm}
\hspace{-0.5cm}
\includegraphics[width=0.5\textwidth]{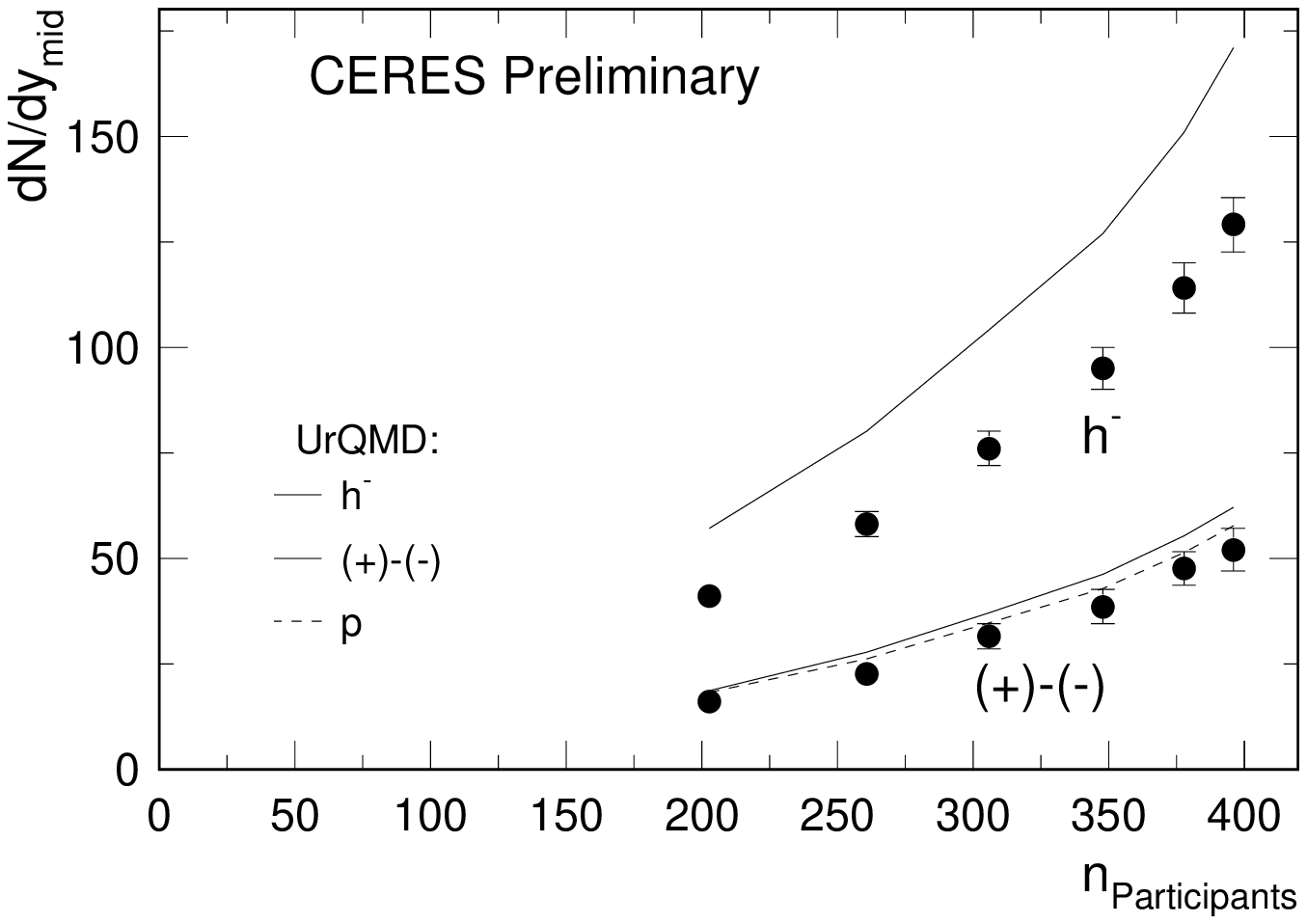}
\caption{Left panel: Measured $h^-$ (left) and proton-like positive net charges (right) transverse mass spectra for events with centrality  $\sigma/\sigma_{\rm geo}$<15\%. Beginning with the lowest rapidity bin the spectra have been multiplied by a successive factor of 10. The lines are the exponential fits to the data. Right panel: Midrapidity yields of $h^-$ and $(+)-(-)$ as a function of number of participants. Also shown are results from UrQMD.}
\label{fig:hadspectra}
\end{figure}

The addition of the TPC allowed for a systematic investigation
of hadronic observables around midrapidity. The matching requirement to
the SIDC system leads to a very efficient suppression of non-vertex
tracks. On the other hand, using SIDC detectors as a veto improves the signal-to-background ratio for the reconstruction of $\Lambda$ and K$^{0}_{s}$. 
The measured transverse mass distributions of negatively charged 
hadrons $h^{-}$ and proton-like positive net charges
`$(+)-(-)$' are shown in Figure~\ref{fig:hadspectra} in different bins of
rapidity for events with centrality $\sigma/\sigma_{\rm geo}$<15\%.
The solid lines represent the best fits to the spectra using an
exponential
$1/m_{t}~{\rm d}N/{\rm d}m_{t}\propto \exp(-m_{t}/T)$.
The rapidity averaged slopes ($T_{h^{-}}=176\pm 5$~MeV and $T_{(+)-(-)}=278\pm 12$~MeV) are similar to those observed at top AGS and SPS energies, 
with only a weak dependence on centrality. The large difference between $h^{-}$ and $(+)-(-)$
slopes 
indicates the presence of strong radial flow also at 40~AGeV. 
 The midrapidity yield of $h^{-}$ as function
of $N_{part}$ (Figure~\ref{fig:hadspectra},~right)
rises significantly stronger than linear,
while a rise with $N_{part}^{1.07\pm0.04}$ at top SPS energy has
been reported recently~\cite{WA98}. This points to a change in 
the particle
production mechanism going from 40~AGeV to 158~AGeV. In
fact, a strongly non-linear behaviour has also been observed 
at the AGS~\cite{Ogli}. 
The non-linear rise of the midrapidity $(+)-(-)$-yield 
with $N_{part}$ is possibly caused by an increasing amount of stopping,
in agreement with the UrQMD prediction.

\begin{figure}[htb]
  \resizebox{.4\textwidth}{!}{\includegraphics{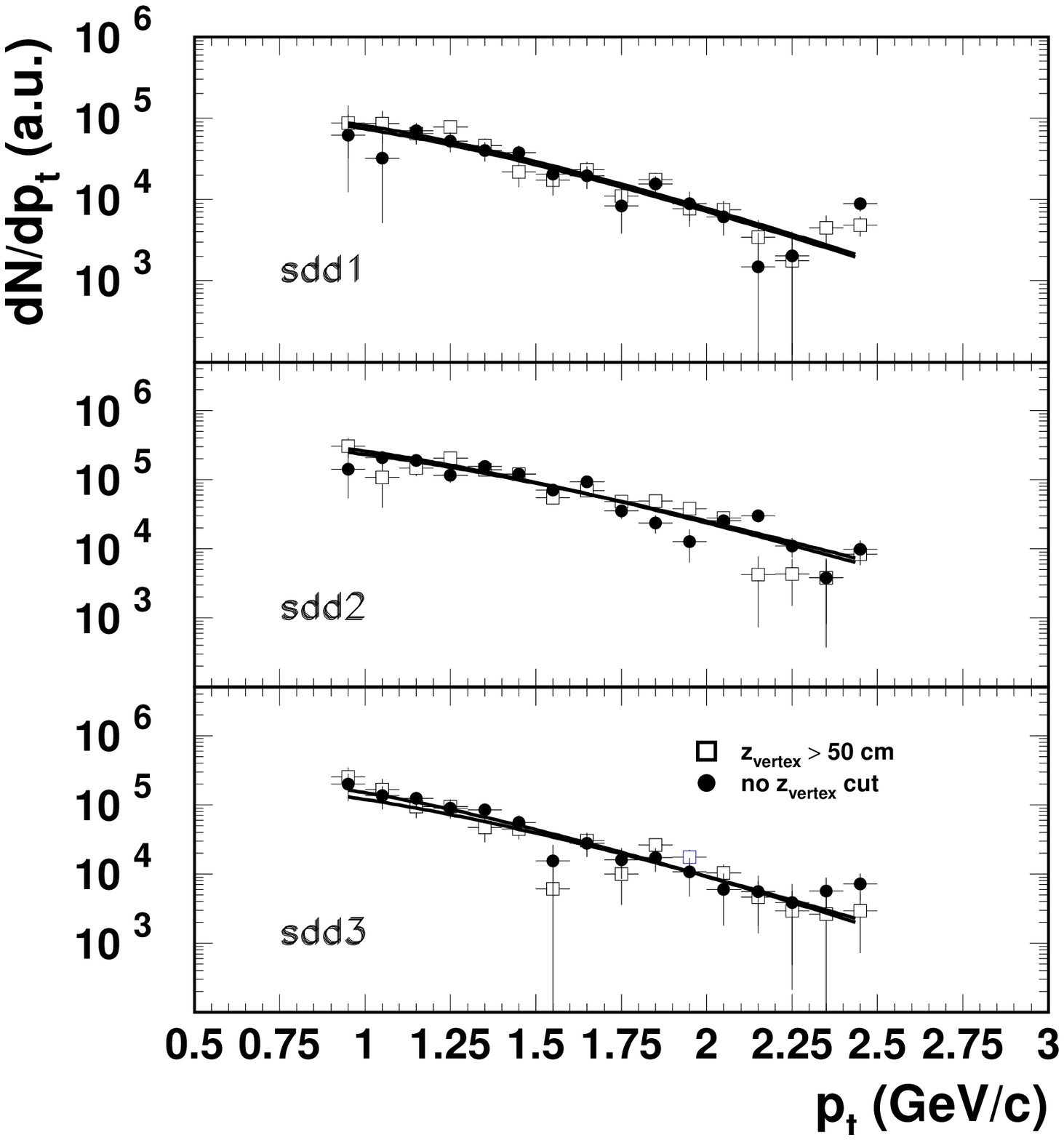}}
\hspace{1.5cm}
\vspace{0.3cm}
  \resizebox{.45\textwidth}{!}{\includegraphics{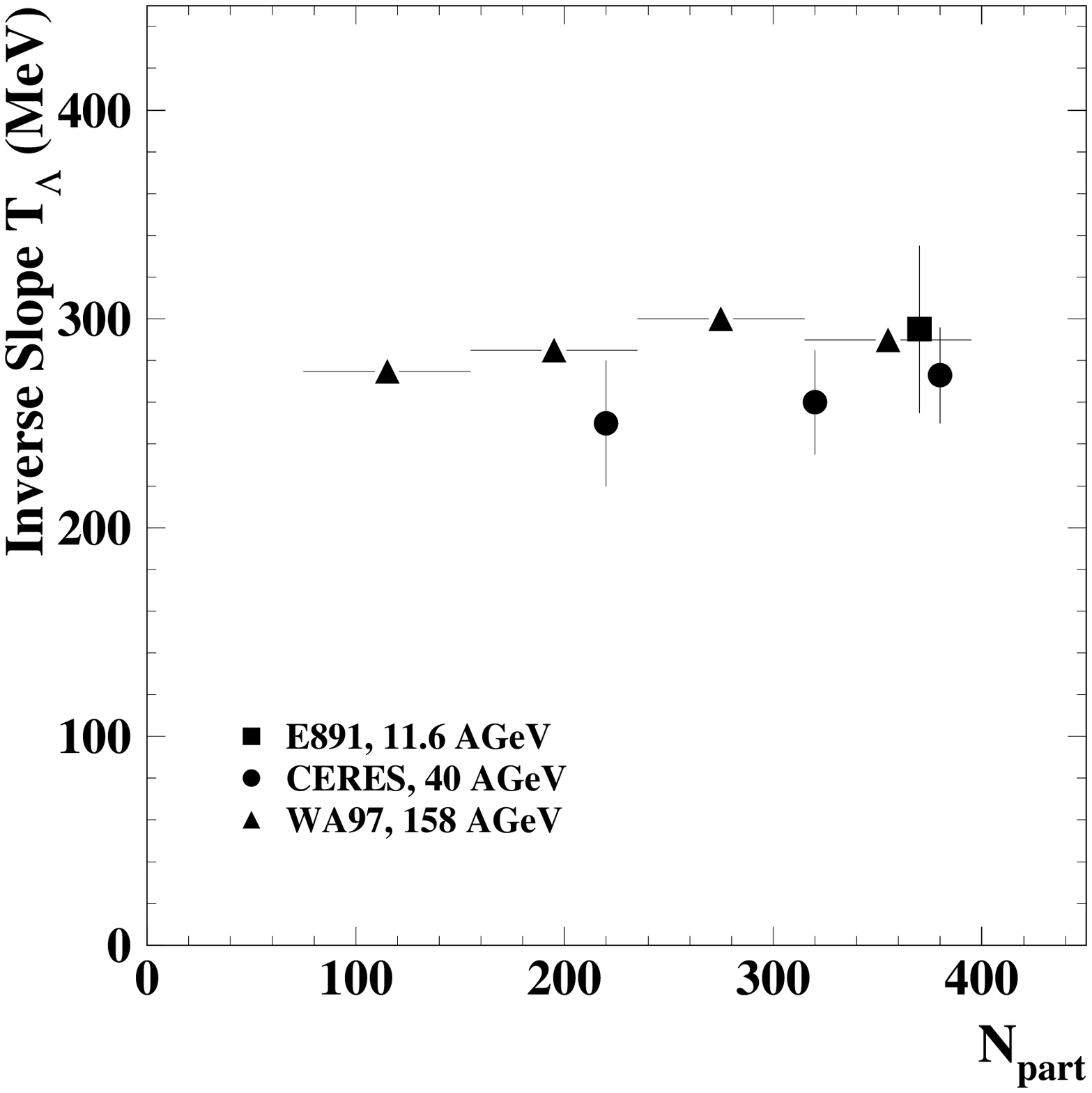}}
  \caption{Left: Measured $\Lambda$ transverse momentum spectra for the three centrality intervals. The solid lines are the exponential fits to the data. Right: $\Lambda$ inverse slope parameters as a function of number of participants.}
\label{fig:lambda}
\end{figure}

We have also identified $\Lambda$ hyperons by invariant mass reconstruction~\cite{wolfgang}. The measurements cover the rapidity range
from $y=$2.0 to 2.4 and transverse momenta from $p_t=0.9$ GeV/c to 2.5 GeV/c.
$\Lambda$ transverse momentum spectra have been fitted by an exponential 
in the three centrality intervals (Figure~\ref{fig:lambda}, left). The extracted inverse slope parameters are shown in
Figure~\ref{fig:lambda} (right). 
The $\Lambda$ slopes measured at 40 AGeV are within errors consistent with those at the top SPS energy \cite{wa97lambda} and AGS \cite{e891lambda}. They are also very similar to the parameters extracted from spectra of proton-like  positive net charges. By integrating the spectra where the data are available and using the results from the exponential fits to extrapolate to infinity and to $p_t$=0 we obtained the yield of $\Lambda$-hyperons in the measured rapidity interval: dN/d$y_\Lambda$ = 11$\pm$2 for centrality $\sigma/\sigma_{\rm geo}$<5\%. We have also measured the ratios of $\Lambda/p=0.22\pm 0.05$ and
$\bar{\Lambda}/\Lambda=0.024\pm 0.010$ for the same centrality.

\section{Directed and Elliptic Flow}

\begin{figure}[htb]
\begin{minipage}[l]{85mm}
\hspace{0.5cm}
\resizebox{9.5cm}{!}{\includegraphics[angle=270]{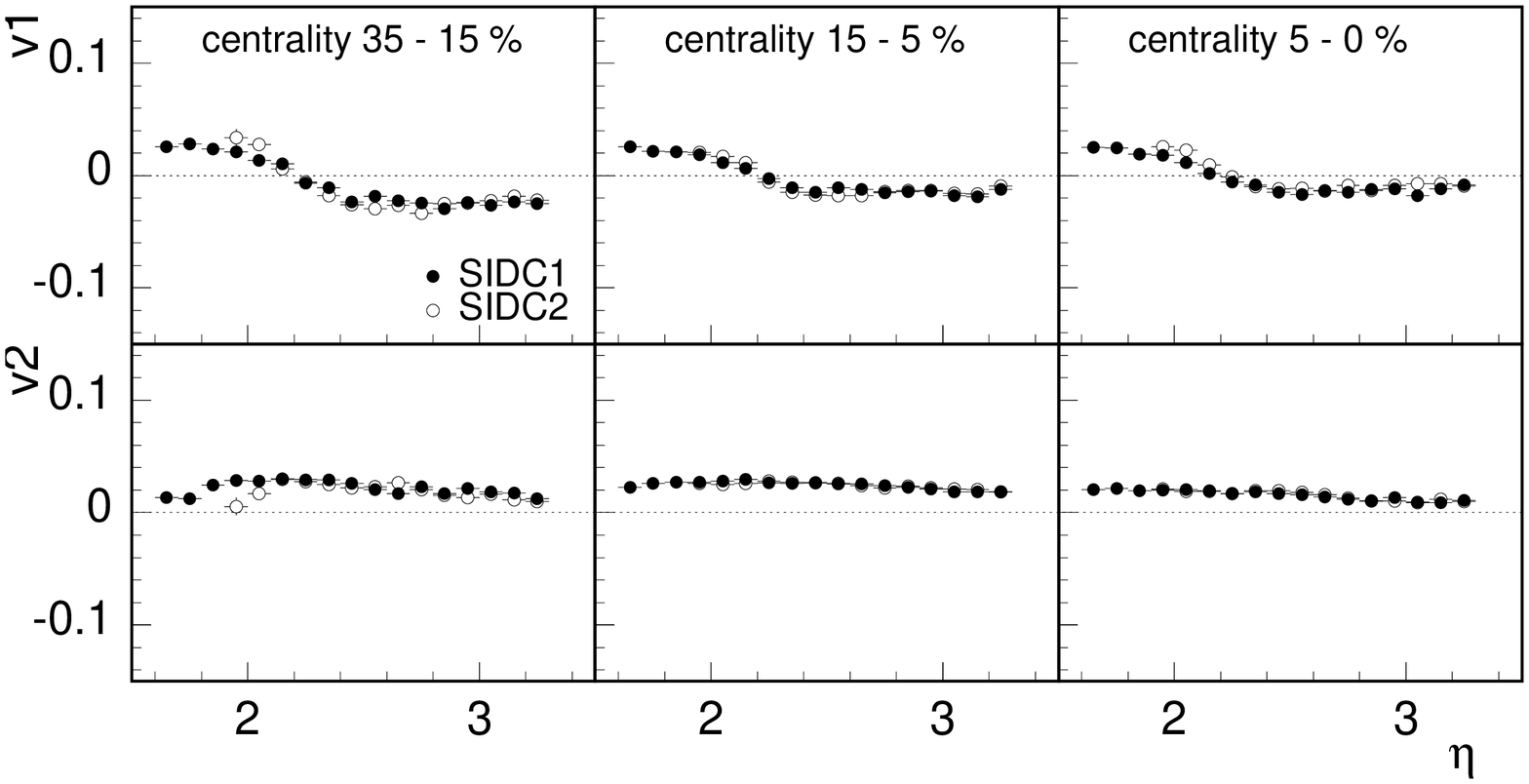}}
\end{minipage}
\begin{minipage}[r]{85mm}
\hspace{1.8cm}
\vspace{-0.5cm}
\includegraphics[width=0.75\textwidth]{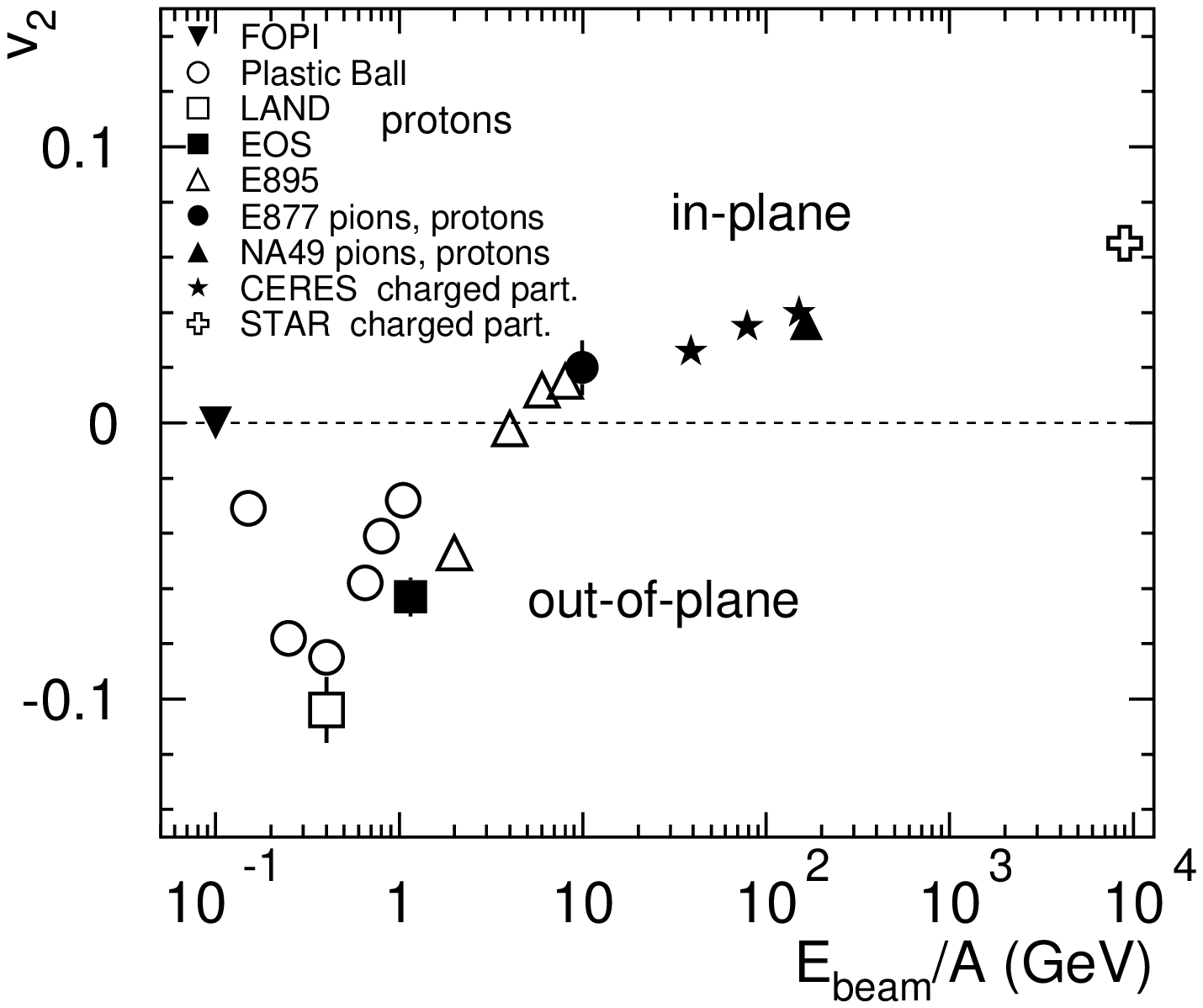}
\vspace{0.3cm}
\end{minipage}
\caption{Left: Pseudorapidity dependence of directed ($v_{1}$) and elliptic ($v_{2}$) flow at 40 AGeV for three centrality intervals. Right: $v_{2}$ as a function of the beam kinetic energy for semi-central collisions of Pb or Au nuclei.}
\label{fig:flow}
\end{figure}

Anisotropies in the azimuthal distribution of particles, also called
anisotropic (directed, elliptic, etc.) transverse flow, have proven
to be sensitive to the 
initial pressure in the collision region and the degree
of thermalization during the expansion stage.
Our flow analysis is based on the azimuthal
hit distributions in the SIDC1,2. 
The two coefficients $v_{1}$ and $v_{2}$, respectively quantifying the
strength of directed and elliptic flow, are obtained from a 
Fourier decomposition of
the azimuthal charged particle distribution with respect to 
the orientation of the reaction plane~\cite{voloflow}. 
The $v_{1}$ and $v_{2}$ values have been corrected for the finite
reaction plane resolution, determined by the
subevent method. The pseudorapidity dependence of directed and elliptic flow 
measured at 40 AGeV is presented in Figure \ref{fig:flow} 
  (left) for three 
different centralities.
While $v_{1}$ exhibits the 
characteristic zero-crossing at midrapidity, $v_{2}$ 
is positive and independent of pseudorapidity. We have also performed the measurements of elliptic flow at 80 and 158 AGeV. The $v_{2}$-values increase with beam energy (Figure~\ref{fig:flow}, right) and fit smoothly into the measured systematics.

\section{Outlook}
 In the fall of 2000 we took a sample of 33M central Pb-Au
events at 158 AGeV with a very good overall performance of the spectrometer.
The analysis of the 2000 data will provide a long-awaited dilepton 
invariant mass spectrum with the mass resolution $\Delta$m/m<2\% in the region of $\omega$ and $\phi$ resonances. The systematic hadron analysis of data taken at 40, 80 and 158 AGeV is in progress, including results on HBT-interferometry and mean transverse momentum fluctuations \cite{harry}.  

\bibliographystyle{aipproc}
\bibliography{my}

\end{document}